# Dielectric, magnetic, and magnetodielectric behaviors of $BaFe_{12}O_{19}$ hexaferrite modulated by Mn and Ti substitutions

Xiao-Fan Zhang[a,b,*], Yang Yang[b,d*], Ze-Qing Guo[a], Li Lv[d], Can Gao[b], Jian-Ping Zhou[a,b,†], Xiao-ming Chen[b]

[a] *Key Laboratory of Functional Materials and Devices for Informatics of Anhui Educational Institutes, Fuyang Normal University, Fuyang, 236037, People's Republic of China*

[b] *School of Physics and Information Technology, Shaanxi Normal University, Xi'an 710119, People's Republic of China*

[c] *School of General Education, Xi'an Mingde Institute of Technology, Xi'an 710124, People's Republic of China*

[d] *Institute of Solid State Physics and Shanxi Province Key Laboratory of Microstructural Electromagnetic Functional Materials, Shanxi Datong University, Datong 037009, People's Republic of China*

**Abstract**: We prepared Mn- and Ti mono-doped and co-doped $BaFe_{12}O_{19}$ hexaferrites via solid-state reaction to investigate the interplay between magnetic and dielectric properties. $Mn^{2+}$ ions preferentially occupy the $4f_2$ and $2b$ sites, while $Ti^{4+}$ ions mainly substitute the $Fe^{3+}$ ions at $4f_1$ and $12k$ sites as revealed by the Raman spectroscopy and formation energy. Pure $BaFe_{12}O_{19}$ exhibits ferrimagnetism. The hexaferrites related to Ti doping have noncollinear longitudinal conical spin order at low temperatures, where $BaFe_6Mn_3Ti_3O_{19}$ retains this spin order up to room temperature. $Ti^{4+}$ substitution at $4f_1$ and $12k$ sites plays a pivotal role in stabilizing the noncollinear conical spin order through adjusting the superexchange interactions and reducing the uniaxial magnetocrystalline anisotropy along the *c*-axis. The magnetic response exhibits two distinct transition temperatures because $Ti^{4+}$ ions interrupt the magnetic superexchange interactions with two inequivalent exchange integrals. Pure $BaFe_{12}O_{19}$ presents a quantum paraelectric behavior at low temperatures, which is disrupted by Mn-Ti co-doping due to the decoupling of electric dipoles within the triangular bipyramid. Electron hopping and polaronic effects dominate the dielectric response at 10 – 50 K,

* Xiao-Fan Zhang and Yang Yang contributed equally to this work.

† Corresponding author. E-mail address: zhoujp@snnu.edu.cn ORCID: 0000-0003-0807-1404, ResearcherID: AGE-2972-2022

while Maxwell-Wagner interfacial polarization and electron hopping contribute to dielectric dispersion at higher temperatures. The negative MD effect of pure $BaFe_{12}O_{19}$ and $BaFe_6Mn_3Ti_3O_{19}$ at 10 K originates from spin-phonon coupling and electric polarization induced by noncollinear spin order under magnetic field, respectively. The Mn-Ti co-doped samples achieve relatively higher MD responses at low magnetic fields. In higher temperatures, the MD effect arises mainly from the magnetic field modulation of the electron hopping with non-intrinsic interfacial polarization. This research reveals the physical properties of M-type hexaferrite can be modulated through substituting $Fe^{3+}$ ions at different sites.

## 1. Introduction

The Magnetodielectric (MD) effect describes the modulation of dielectric properties by an external magnetic field. Many mechanisms, including spin-phonon coupling, magnetoelectric effect, electron hopping, and interface polarization, can induce this phenomenon [1-4]. The research on MD effect is vital to understanding the fundamental interactions between spin, electrons, orbitals, phonons, etc [5]. Moreover, it enables the development of next-generation magnetoelectric memory devices with multifunctional capabilities and miniaturized components [6,7]. M-type hexaferrite with magnetoplumbite structure represents an important class of type II single-phase multiferroic materials [3]. In these systems, ion doping can induce noncollinear magnetic structures that enable electric polarization through magnetic field excitation, demonstrating intrinsic magnetoelectric coupling and consequently producing intrinsic MD effects. At low temperatures, additional intrinsic MD effects may emerge from spin-phonon coupling and magnetic-field-modulated electron hopping [8]. As the temperature increases, extrinsic MD effects result from the combination of magnetoresistance and Maxwell-Wagner interfacial polarization [9].

M-type hexaferrites are dielectric materials with significantly higher resistivity than the conventional magnetic metals. This characteristic, accompanied by low eddy current losses, makes them ideal permanent magnet materials for small generators and motors. Their large uniaxial magnetocrystalline anisotropy along the *c*-axis can be further applied in magnetic recording devices. M-type hexaferrite ($AFe_{12}O_{19}$: $A$ = Ba, Sr, Pb, etc.) is formed by $S$ ($2Fe_3O_4$), $R$ ($BaFe_6O_{11}$), $S$*, and $R$* ($S$ and $R$ rotated by 180° around the *c*-axis) blocks stacked along the [0 0 1] crystallographic direction in an *SRS**R** sequence as shown in Fig. 1(a) [10-12]. Inside the crystal cell, $Fe^{3+}$ ions are located in three distinct oxygen polyhedra: octahedral $FeO_6$, tetrahedral $FeO_4$, and triangular bipyramidal $FeO_5$. The interstitial sites of oxygen octahedral $FeO_6$ and tetrahedral $FeO_4$ in $S$ block are $2a$ and $4f_1$ sites, respectively, those of octahedral $FeO_6$ and triangular bipyramidal $FeO_5$ in $R$ block are $4f_2$ and $2b$ sites, respectively, and that of octahedral $FeO_6$ at the junction of $S$ and $R$ blocks is $12k$ site. The $Fe^{3+}$ ions within different oxygen octahedra have different spin directions. The magnetic structure arises from the antiparallel alignment of $Fe^{3+}$ ions along the *c*-axis, with spin-up moments at $2a$, $2b$, and $12k$ sites and spin-down moments at $4f_1$ and $4f_2$ sites [10]. The M-type hexaferrite contains 24 $Fe^{3+}$ ions in a unit cell, distributed as $Fe(12k):Fe(2a):Fe(2b):Fe(4f_1):Fe(4f_2) = 6:1:1:2:2$. It presents ferrimagnetic characteristics because of the different $Fe^{3+}$ ions in the spin-up and spin-down directions. The magnetic properties are primarily governed by four key superexchange interactions mediated $O^{2-}$ ions, i.e., $Fe^{3+}(2b)$-$O^{2-}$-$Fe^{3+}(4f_2)$, $Fe^{3+}(12k)$-$O^{2-}$-$Fe^{3+}(4f_2)$, $Fe^{3+}(12k)$-$O^{2-}$-$Fe^{3+}(4f_1)$ and $Fe^{3+}(2a)$-$O^{2-}$-$Fe^{3+}(4f_1)$ interactions under consideration of $Fe^{3+}$ spins as shown in Fig. 1(b). The noncollinear conical spin ordering formed through the interaction between these magnetic ions propagates along the *c*-axis to produce strong uniaxial magnetocrystalline anisotropy.

The substitution will change the superexchange interactions between $Fe^{3+}$ ions and reduce the uniaxial magnetocrystalline anisotropy, as a result, inducing the noncollinear magnetic structure and electric dipoles. Appropriate substitution of $Fe^{3+}$ ions in the oxygen polyhedra with elements, such as Sc [10,13-16], In [8,17], Ti [18], Cu-Ti [19],

Co-Ti [20-22], Co-La [23] and Mg-Ir [24], etc. can stabilize the ferroelectric state to room temperature with high magnetoelectric and MD couplings [11]. In addition, the $Fe^{3+}$ ion at 2*b* site deviates from the symmetric center of the $FeO_5$ bipyramid, generating an intrinsic electric dipole [8,25].

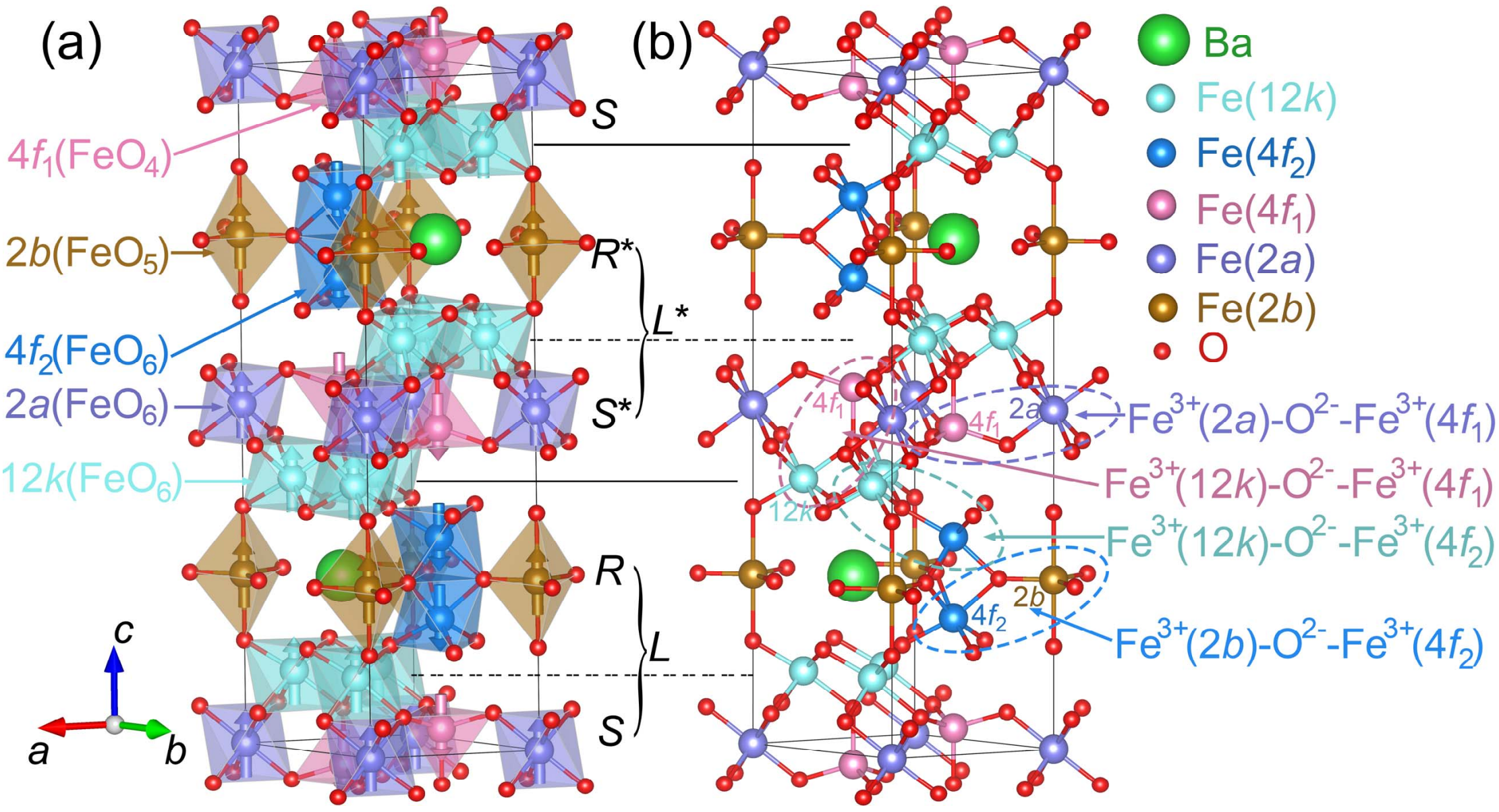

Fig. 1 Schematic of $BaFe_{12}O_{19}$ crystal structure with (a) polyhedron with spin direction of Fe ions and (b) ball-and-stick modeling with main superexchange interaction. $S$, $S^*$, $R$, and $R^*$ blocks are the basic units of the crystal structure. $L$ and $L^*$ blocks are the basic units of the magnetic structure.

Co-Ti co-doping can induce the noncollinear magnetic structures in M-type hexaferrite $PbCoTiFe_{10}O_{19}$ and keep the noncollinear spin order up to 240 K [20]. However, the distinct roles of $Co^{3+}$ and $Ti^{4+}$ ions in modulating these magnetic structures remain unclear, which hinders the fundamental understanding of noncollinear spin order and the rational design of new materials with tailored magnetic properties. In this article, we present a systematic investigation of Mn-Ti co-doped, Mn and Ti mono-doped $BaFe_{12}O_{19}$ hexaferrites, revealing the superexchange interaction between the iron ions at different sites in Mn-Ti co-doped samples. The noncollinear spin order in $BaFe_6Mn_3Ti_3O_{19}$ can be maintained up to room temperature. Then, we investigated the effect of altered superexchange interactions between ions on the dielectric and MD properties.

## 2. Experimental details

### 2.1. Sample preparation

A series of M-type hexaferrite ceramics with nominal compositions $BaFe_{12-2x}Mn_xTi_xO_{19}$ ($x_{Mn\text{-}Ti}$ = 0, 1, 2, and 3), $BaFe_{12-x}Mn_xO_{19}$ ($x_{Mn}$ = 0, 1, 2, and 3), and $BaFe_{12-x}Ti_xO_{19}$ ($x_{Ti}$ = 0.0, 0.5, 1.0, and 1.5) were synthesized via the conventional solid-state reaction method. High-purity (99.9%) starting materials, including $BaCO_3$, $Fe_2O_3$, $MnO_2$, and $TiO_2$, were precisely weighed according to the stoichiometric ratios and thoroughly mixed with a planetary ball mill for homogeneous distribution. The mixed powders were initially sintered in air to initiate solid-state reactions. The resulting product was subsequently reground to ensure phase uniformity and then uniaxially pressed into tablets (about 7.0 mm in diameter and 0.8 − 1.2 mm in thickness) under optimized compaction pressure. Finally, the pressed tablets were calcined in an oxygen atmosphere (1 atm) and annealed for 10 h at relatively low temperatures in an oxygen atmosphere (1 atm) to ensure proper oxygen stoichiometry, phase stability and appropriate grain size [8,26,27]. Detailed thermal treatment parameters for each composition are summarized in Table 1.

Table 1 Main preparation conditions of ceramics.

| Chemical Formula | Pre-sintering | Calcining | Annealing |
|---|---|---|---|
| $BaFe_{12-2x}Mn_xTi_xO_{19}$ ($x_{Mn\text{-}Ti}$ = 0, 1, 2, and 3) | Atmosphere 1100 °C, 15 h | 1 atm $O_2$ 1150 °C, 15 h | 1 atm $O_2$ 1000 °C, 10 h |
| $BaFe_{12-x}Mn_xO_{19}$ ($x_{Mn}$ = 0, 1, 2, and 3) | Atmosphere 1000 °C, 10 h | 1 atm $O_2$ 1100 °C, 15 h | 1 atm $O_2$ 1000 °C, 10 h |
| $BaFe_{12-x}Ti_xO_{19}$ ($x_{Ti}$ = 0.0, 0.5, 1.0, and 1.5) | Atmosphere 1150 °C, 15 h | 1 atm $O_2$ 1200 °C, 15 h | 1 atm $O_2$ 1000 °C, 10 h |

### 2.2. Measurements

The crystal structure of samples was characterized by X-ray diffraction (XRD) using a Rigaku D/Max 2550 diffractometer with Cu Kα radiation. The Raman spectra were collected on a spectrometer (Horiba Jobin Yvon, LabRAM HR Evolution) equipped with a 532 nm laser source. The magnetization dependence on temperature and magnetic field was measured by a multifunctional physical measurement system (Quantum Design, PPMS-9) under both zero-field-cooled (ZFC) and field-cooled (FC) conditions. The MD response was characterized by an MD testing system. In this system, the temperature and magnetic field were controlled by a temperature controller (East Changing Technologies, TC280) and a superconducting magnet (Cryogenic, CFM-6T-150-RT), respectively. The dielectric permittivity was measured by an impedance analyzer (Keysight, E4980AL). In addition, the temperature dependence of dielectric permittivity was tested at 10 Hz–100 kHz from 10 K to 300 K by the MD testing system and at 1 Hz–100 kHz from 150 K to 330 K by a broadband dielectric spectroscopy (Novocontrol, Alpha-A).

### 2.3. Computational calculation

First-principles calculations were performed to determine the formation energies of doped configurations using the Vienna ab initio simulation package (VASP) [28] in MedeA® materials modeling environment [29]. Core electrons were treated using the projector-augmented wave (PAW) method [30], while valence electrons were expanded in a plane-wave basis set with a kinetic energy cutoff of 400 eV. The Brillouin zone integration was performed using a Γ-centered k-point mesh with a spacing of 0.25/Å. Structural optimizations were carried out until the forces on all atoms were below 0.02 eV/Å, ensuring accurate atomic positions. The electronic self-consistent iteration process was considered converged when the total energy difference between successive steps was less than $1 \times 10^{-5}$ eV.

## 3. Results and discussion

### 3.1. Crystal structure

Fig. 2(a) presents the XRD patterns of $BaFe_{12-2x}Mn_xTi_xO_{19}$ ($x_{Mn\text{-}Ti}$ = 0, 1, 2, and 3) ceramics. The diffraction peaks of pure $BaFe_{12}O_{19}$ sample and Mn-Ti co-doped samples with $x_{Mn\text{-}Ti}$ = 1 and 2 well match the standard M-type hexaferrite structure (PDF No. 84-0757, space group: $P6_3/mmc$), indicating successful synthesis of pure-phase hexaferrite. A small amount of secondary phase $Ba_{1.12}Ti_8O_{16}$ (PDF No. 77-0883, space group: I4) appears in the heavily doped sample ($x_{Mn\text{-}Ti}$ = 3), attributable to exceeding the solubility limit of $Ti^{4+}$ ions. The XRD patterns of $BaFe_{12-x}Mn_xO_{19}$ ($x_{Mn}$ = 0, 1, 2, and 3) and $BaFe_{12-x}Ti_xO_{19}$ ($x_{Ti}$ = 0.0, 0.5, 1.0, and 1.5) ceramics are shown in Fig. 2(b) and 2(c) for comparison. The Mn mono-doped samples maintain a pure hexaferrite structure. But foreign phases appear in the Ti mono-doped sample with a small amount of $x_{Ti}$ = 1.5. Thus, Mn-Ti co-doping could increase the Ti content in $BaFe_{12}O_{19}$ lattice. The lattice parameters $a$, $c$, and unit cell volume fitted by Jade 6.0 are very close after doping $Mn^{2+}$ ions with the same radius (0.66 Å) and slightly decrease after doping $Ti^{4+}$ ions with a smaller radius (0.605 Å) in comparison with the $Fe^{3+}$ cation radius (0.64 Å). This contraction is particularly evident in the Mn-Ti co-doped samples as shown in Fig. 2.

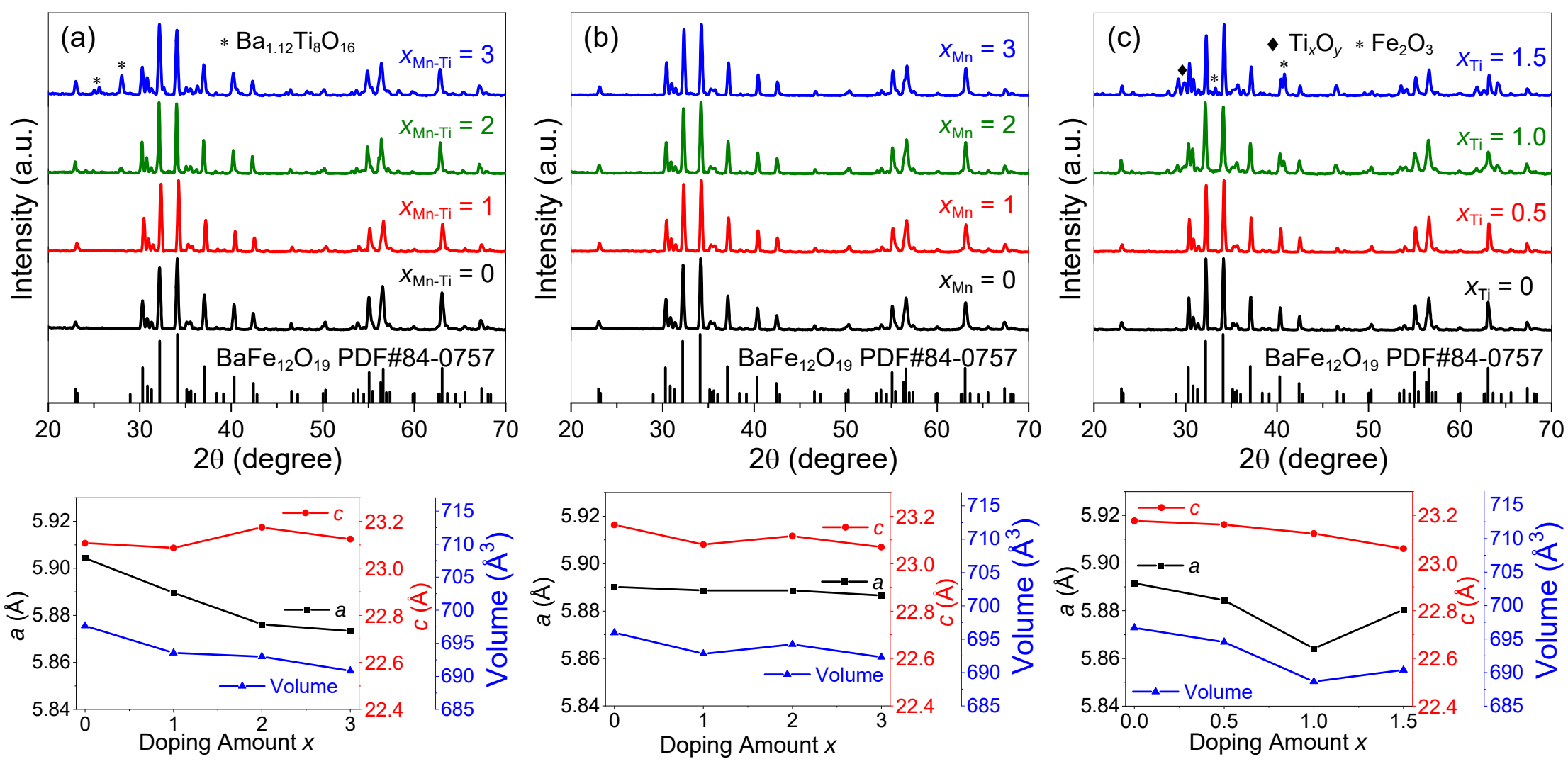

Fig. 2 XRD patterns, lattice parameters, and unit cell volumes of (a) $BaFe_{12-2x}Mn_xTi_xO_{19}$ ($x$ = 0, 1, 2, and 3), (b) $BaFe_{12-x}Mn_xO_{19}$ ($x$ = 0, 1, 2, and 3), and (c) $BaFe_{12-x}Ti_xO_{19}$ ($x$ = 0, 0.5, 1.0, and 1.5).

The M-type hexaferrite structure contains $Fe^{3+}$ ions distributed across five crystallographically distinct sites within the unit cell, giving rise to four principal superexchange interactions as shown in Fig. 1. The Mn-Ti substitution for $Fe^{3+}$ ions significantly modifies the Fe-O vibrational modes, as evidenced by characteristic changes in the Raman spectra. Fig. 3(a) displays the Raman spectra of $BaFe_{12-2x}Mn_xTi_xO_{19}$ ($x_{Mn-Ti}$ = 0, 1, 2, and 3) samples in the 300 − 800 $cm^{-1}$ range. The Lorenz-Gaussian function was used to determine these peaks. There are seven well-defined vibrational modes in the pure $BaFe_{12}O_{19}$ at around 330.2, 413.8, 470.3, 529.8, 614.9, 687.2, and 723.9 $cm^{-1}$, labeled as peaks I, II, III, IV, V, VI, and VII, respectively.

Peak I is relative to the Fe-O bonding vibrations inside all octahedra. Peak II originates from the Fe-O vibrations inside the octahedra at the junction of *R* and *S* blocks. Peak III comes from Fe-O vibrations in the octahedra in *S* block and the octahedra at the junction position of *R* and *S* blocks. Peaks V, VI, and VII are from the Fe-O vibrations inside the octahedra in *R* block, the bipyramids in *R* block, and the tetrahedra in *S* block, respectively [15,31-33].

The Raman spectra of the Mn-Ti co-doped samples exhibit significant modifications compared to the pure sample. The positions of peaks V and VII gradually shift to higher wavelengths with the increase in Mn-Ti amount, and the intensity of peak V is strengthened. The positions of peaks I − VI also changed obviously with the increase in doping amount. These indicate the Mn-Ti ions mainly substitute for the $Fe^{3+}$ ions in the oxygen octahedral $FeO_6$ ($4f_2$ site) in *R* block, the oxygen bipyramidal $FeO_5(2b)$ in *R* block, the oxygen tetrahedral $FeO_4$ ($4f_1$ site) in *S* block, and the oxygen octahedral $FeO_6$ ($12k$) at the junction.

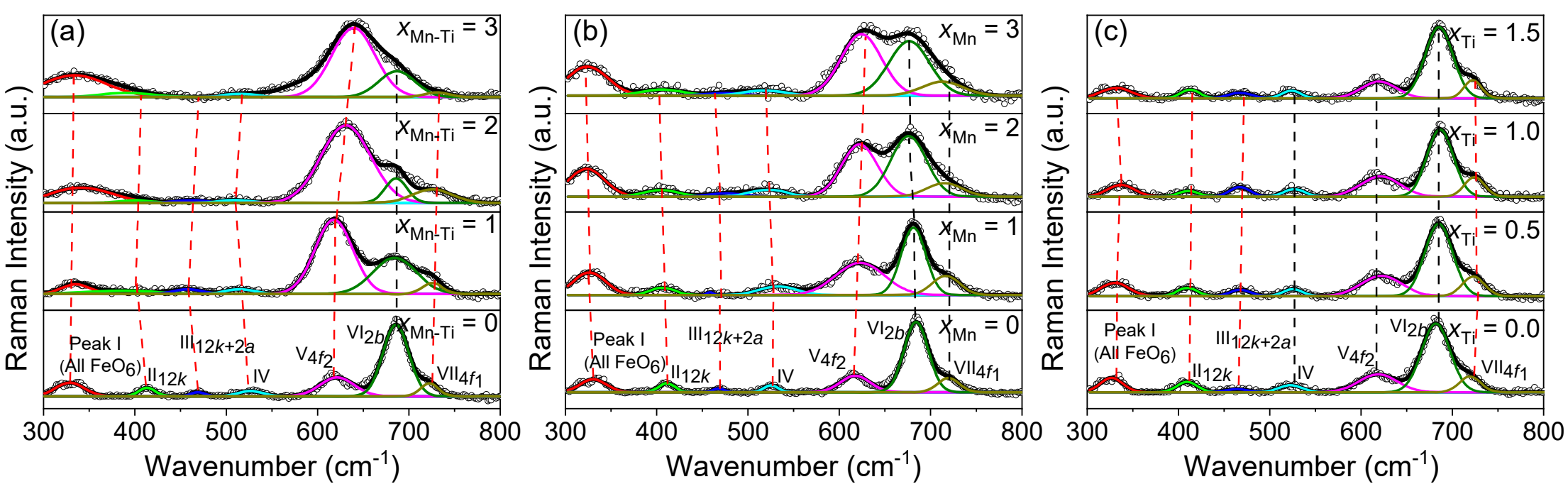

Fig. 3 Raman spectra of (a) $BaFe_{12-2x}Mn_xTi_xO_{19}$ ($x$ = 0, 1, 2, and 3), (b) $BaFe_{12-x}Mn_xO_{19}$ ($x$ = 0, 1, 2, and 3), and (c) $BaFe_{12-x}Ti_xO_{19}$ ($x$ = 0, 0.5, 1.0, and 1.5) ceramics.

We measured Raman spectra of $BaFe_{12-x}Mn_xO_{19}$ ($x_{Mn}$ = 0, 1, 2, and 3) and $BaFe_{12-x}Ti_xO_{19}$ ($x_{Ti}$ = 0.0, 0.5, 1.0, and 1.5) samples to investigate the occupancies of $Mn^{2+}$ and $Ti^{4+}$ ions. $Mn^{2+}$ substitution induces a pronounced blueshift and significant intensity enhancement of Peak V, as shown in Fig. 3(b). Peaks I-IV exhibit minor modifications, and Peak VI occurs a little blueshift with enhanced full width at half maximum. These spectral changes demonstrate that $Mn^{2+}$ ions preferentially substitute for $Fe^{3+}$ ions at the oxygen octahedral $4f_2$ and oxygen bipyramidal $2b$ sites within the *R* block. Whereas Peaks III and VII show marked blueshifts and intensity increases, accompanied by a slight shift in Peak II for the Ti mono-doped samples with $x_{Ti}$ = 0.5 and 1.0, as shown in Fig. 3(c). These characteristics in the vibrational spectra suggest that $Ti^{4+}$ ions predominantly occupy both the tetrahedral $4f_1$ site in *S* block and the octahedral $12k$ site at the junction between *R* and *S* blocks [18,31].

We calculated the formation energy of Mn and Ti mono-doped $BaFe_{12}O_{19}$ hexaferrites to verify the preferential occupation of Mn and Ti dopants by density functional theory. The formation energy of doping materials is expressed as [34,35]

$$E_{form} = E_{dope} - E_{bulk} - n\mu_{\mathrm{M}} + n\mu_{N} \tag{1}$$

where $E_{dope}$ and $E_{bulk}$ represent the total energy after and before doping atoms, respectively. $\mu_{\mathrm{M}}$ and $\mu_{\mathrm{N}}$ denote the chemical potentials of dopant and substituted host element, respectively. *n* is the number of substitutions. The chemical potential $\mu_{\mathrm{M}}$ was determined from its corresponding stable metal oxide by

$$\mu_{\mathrm{M}} = (\mu_{\mathrm{M}_a\mathrm{O}_b} - b\mu_{\mathrm{O}}) / a \tag{2}$$

where $\mu_{\mathrm{M}_a\mathrm{O}_b}$ is the chemical potential of the metal oxide and $\mu_{\mathrm{O}} = \mu_{\mathrm{O}_2} / 2$ is the oxygen chemical potential. Then, we get $\mu_{\mathrm{Fe}} = -11.932$ eV, $\mu_{\mathrm{Mn}} = -10.496$ eV, and $\mu_{\mathrm{Ti}} = -11.837$ eV.

Fig. 4 shows the calculated formation energies for Mn and Ti substitutions at different Fe sites in $BaFe_{12}O_{19}$. The negative formation energies for both dopants indicate favorable incorporation of Mn and Ti into the hexaferrite lattice. Mn substitution is energetically preferred at the $2b$ and $4f_2$ sites, while Ti shows greater stability at the $4f_1$ and $12k$ sites, agreeing with the Raman results.

Briefly, the $Mn^{2+}$ ions prefer the $4f_2$ and $2b$ sites while $Ti^{4+}$ ions tend to enter the $4f_1$ and $12k$ sites. This site-selective doping behavior has significant implications for the magnetic properties. The substitution could modulate the superexchange interactions among $Fe^{3+}$ ions. Among them, the replacement of $Fe^{3+}$ ions in $12k$ site will play an important role in destroying the uniaxial magnetocrystalline anisotropy along the *c*-axis and inducing the noncollinear magnetic structures because these $Fe^{3+}$ ions serve as a crucial bridge connecting the magnetic sublattices in $BaFe_{12}O_{19}$.

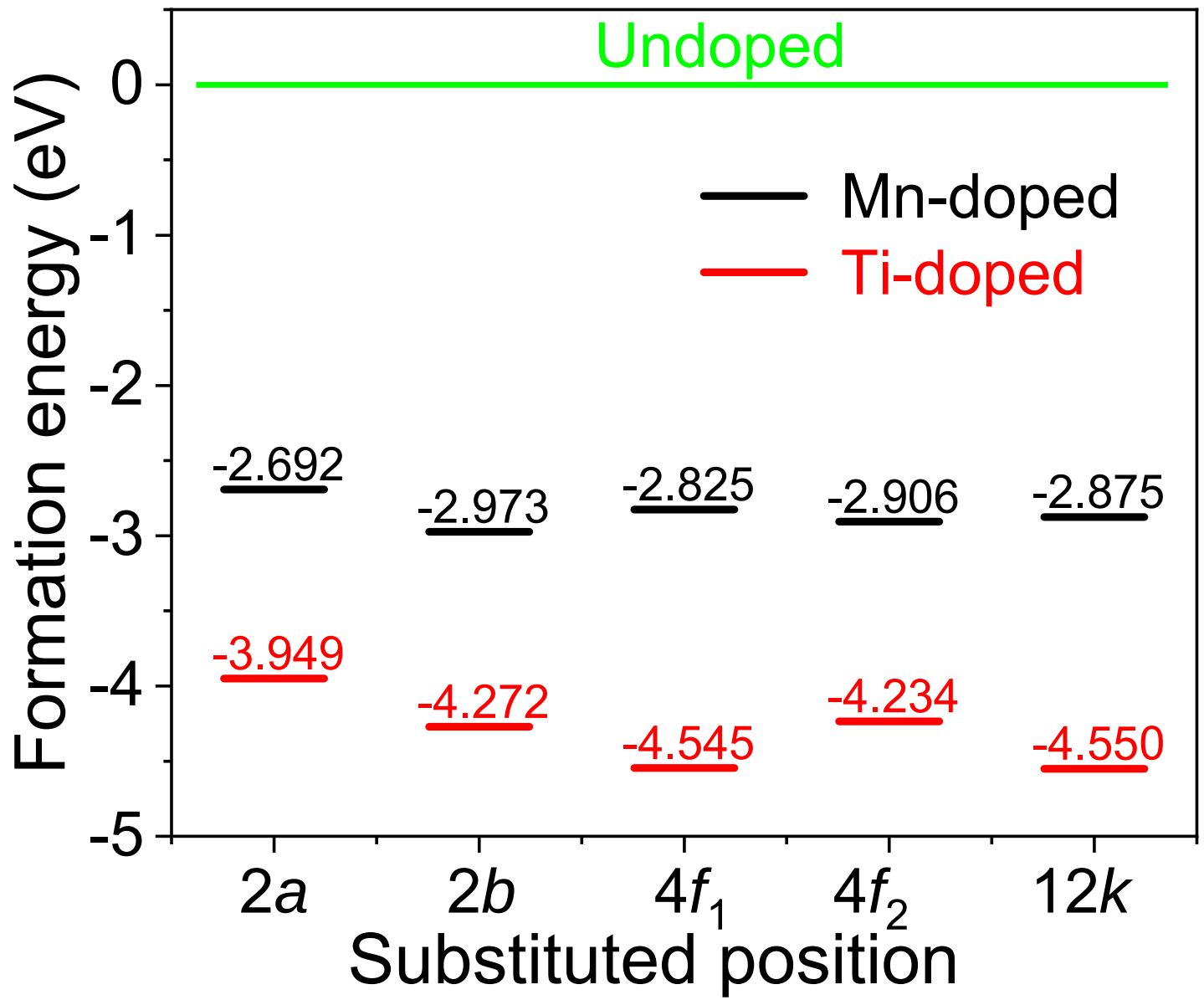

Fig. 4 Calculated formation energy for Mn or Ti substitutions at different Fe sites in $Ba_2Fe_{24}O_{38}$ unit cell.

**3.2. Magnetic property**

Fig. 5(a)–5(d) display the magnetization with temperature for the $BaFe_{12-2x}Mn_xTi_xO_{19}$ ($x_{Mn\text{-}Ti}$ = 0, 1, 2, and 3) ceramics measured under ZFC and FC processes in the 5 − 300 K range. The pure $BaFe_{12}O_{19}$ exhibits characteristic collinear ferrimagnetism with minimal temperature dependence, as confirmed by the near-zero ZFC derivative. The ferrimagnetism arises from the net magnetic moment difference between spin-up and spin-down sublattices. The distinct temperature evolution of magnetization at different sublattices produces the characteristic peak in the ZFC curve. The decrease in FC magnetization at high temperatures is from the thermal weakening of superexchange interactions [8,25].

The Mn-Ti co-doped samples exhibit significantly different ZFC characteristics in comparison with pure $BaFe_{12}O_{19}$, as shown in Fig. 5(b) − 5(d). The ZFC magnetization of samples with $x_{Mn\text{-}Ti}$ = 1 and 2 varies significantly with temperature, showing peaks at $T_{M2}$ of 135 K and 242 K, respectively, attributed to the transition from noncollinear longitudinal conical spin order to collinear ferrimagnetism (Fig. 7) [8,10,20-22,36,37]. The $T_{M2}$ is extracted from the first derivative (at d$M$/d$T$ = 0) of the respective ZFC data. In contrast, the sample with $x_{\text{Mn-Ti}}$ = 3 increases monotonously, indicating its noncollinear spin order expands over room temperature. The apex angle of spin-cone decreases with increasing temperature, which can cause the rotation of magnetic moment from the direction away from the $c$-axis to alignment along the $c$-axis as shown in the inset in Fig. 5(c), leading to the increase in ZFC magnetization with temperature in noncollinear magnetic phase [19,24]. Furthermore, $T_{M2}$ shifts to higher temperatures with Mn-Ti doping amount, indicating the enhancement of noncollinear spin order. The FC magnetization of Mn-Ti co-doped samples with $x_{\text{Mn-Ti}}$ = 1 and 2 does not change significantly in 5 K − $T_{M2}$, as shown in Fig. 5(b) and 5(c).

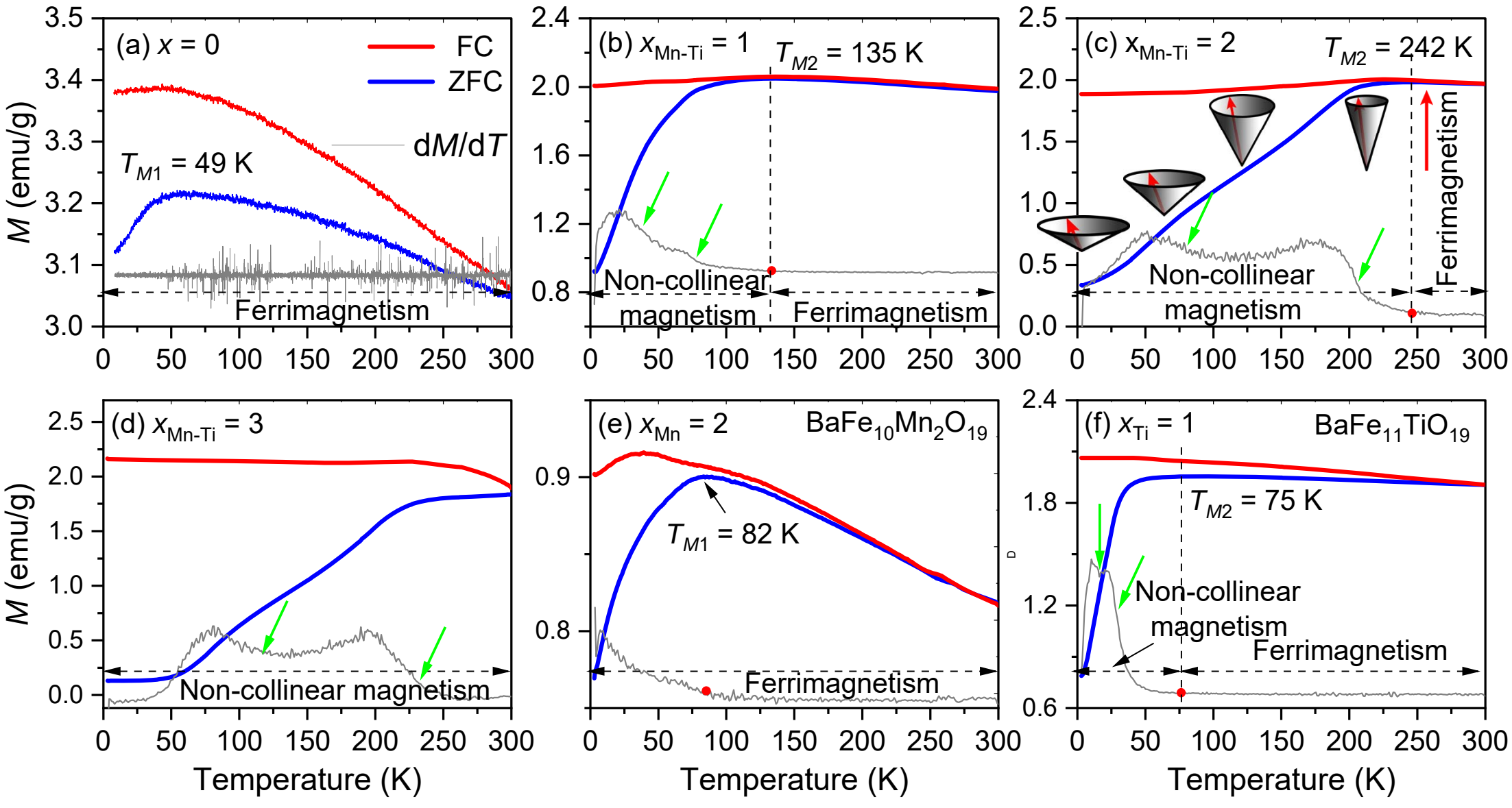

Fig. 5 Temperature dependence of the zero field-cooling (ZFC), field-cooling (FC) magnetization and first derivative d$M$/d$T$ of ZFC for (a)−(d) $BaFe_{12-2x}Mn_xTi_xO_{19}$ ($x$ = 0, 1, 2, and 3), (e) $BaFe_{10}Mn_2O_{19}$, and (f) $BaFe_{11}TiO_{19}$ ceramics. The inset in (c) shows the spin-cone transition with temperature. The red dot represents the transition from positive d$M$/d$T$ to negative d$M$/d$T$. The testing field is 100 Oe, and the field-cooling (FC) field is also 100 Oe.

The FC magnetization of the sample with $x_{\text{Mn-Ti}}$ = 3 decreases monotonously in 5−300 K as shown in Fig. 5(d). The doped single crystal M-type hexaferrite exhibits the noncollinear spin order when the external magnetic field is parallel to the [0 0 1] crystal orientation, while exhibits collinear ferrimagnetism when the magnetic field is perpendicular to the [0 0 1] crystal direction [14]. The magnetism in polycrystalline ceramic is an average over all crystal orientations, so its FC magnetization will decrease monotonously with temperature since the FC cooling process could enhance the ferrimagnetic contribution [33]. The significant divergence between the ZFC and FC curves confirms that the Mn-Ti co-doped M-type hexaferrites maintain noncollinear spin order along some specific crystallographic directions[14].

The temperature dependence of magnetization for the $BaFe_{10}Mn_2O_{19}$ and $BaFe_{11}TiO_{19}$ is shown in Fig. 5(e) and 5(f), respectively, revealing the distinct dopant effects on the M-type hexaferrite magnetism. $BaFe_{10}Mn_2O_{19}$ maintains collinear ferrimagnetism like the pure $BaFe_{12}O_{19}$ with minimal temperature variation. In contrast, $BaFe_{11}TiO_{19}$ shows a clear ZFC peak at $T_{M2}$, mirroring the behavior of $BaFe_{12-2x}Mn_xTi_xO_{19}$ with noncollinear spin order in 5 K − $T_{M2}$. On the other hand, Mn-Ti co-doped and Ti mono-doped samples enjoy a significant increase in magnetization from 10 K to $T_{M2}$ in comparison with the $BaFe_{12}O_{19}$ and Mn mono-doped samples as listed in Table 2. This demonstrates that $Ti^{4+}$ ions, rather than $Mn^{2+}$ ions, primarily drive the formation of noncollinear spin order below $T_{M2}$.

Table 2 Change in magnetization for the samples from 8 K to $T_M$ under the ZFC process.

| Samples | $BaFe_{12}O_{19}$ | $x_{Mn\text{-}Ti} = 1$ | $x_{Mn\text{-}Ti} = 2$ | $x_{Mn\text{-}Ti} = 3$ | $x_{Mn} = 2$ | $x_{Ti} = 1$ |
|---|---|---|---|---|---|---|
| At 8 K (emu/g) | 3.125 | 0.968 | 0.336 | 0.131 | 0.786 | 0.856 |
| At $T_{M2}$ or $T_{M1}$ (emu/g) | 3.212 | 2.048 | 1.983 | 1.838 | 0.899 | 1.952 |
| Rate (%) | 2.784 | 111.6 | 490.18 | 1303.1 | 14.38 | 128.04 |

Another distinctive feature is the appearance of two inflection points in the FC curves for both Ti mono-doped and Mn-Ti co-doped samples. These are clearly evidenced by two peaks in their first derivatives as shown in Fig. 5(b)–5(d) and 5(f). The two inflection points also appear on the ZFC curves of M-type hexaferrites related to Ti mono-doping [18,19,21], suggesting this characteristic is specifically associated with Ti mono-doping.

Fig. 6(a) − 6(e) display the hysteresis loops of $BaFe_{12-2x}Mn_xTi_xO_{19}$ ($x_{Mn\text{-}Ti}$ = 0, 2, and 3), $BaFe_{10}Mn_2O_{19}$, and $BaFe_{11}TiO_{19}$ ceramics measured at some temperatures. Their initial magnetization curves were first measured at 10 K to reveal the magnetic behaviors as shown in the insets. There is a linear relationship with low slope between magnetization and magnetic field in the initial magnetization curves for the Mn-Ti co-doped $BaFe_{12-2x}Mn_xTi_x$O19 and $BaFe_{11}TiO_{19}$ samples under a low magnetic field at 10 K as marked by green circles in Fig. 6(b), 6(c) and 6(e), corresponding to the alteration of canting angle of spin-cone within the external magnetic field. Consequently, the initial magnetization curves of Mn-Ti co-doped $BaFe_{12-2x}Mn_xTi_xO_{19}$ samples extend partially outside their hysteresis loops, which is a characteristic feature of noncollinear spin order [3,38-40]. The initial magnetization curve of $BaFe_{11}TiO_{19}$ is very close to the right branch of hysteresis loop due to its low Ti content. In contrast, the initial magnetization curves remain entirely within the hysteresis loops for the pure $BaFe_{12}O_{19}$ and $BaFe_{10}Mn_2O_{19}$, representing collinear ferrimagnetic behavior. These magnetic characteristics agree with the thermomagnetic results in Fig. 5.

The magnetic structure evolves under applied magnetic fields through several mechanisms in these noncollinear systems. The evolution of noncollinear spin order, involving the spin-cone axis, the direction of magnetic moment of basic unit, and the spin-cone angle, is usually different between the initial magnetization stage of 0 − 50 kOe and the magnetization stage of ±50 kOe, resulting in the deviation of the initial magnetization curve from the main hysteresis loop.

Fig. 6(f) summarizes the coercivity and saturation magnetization of all samples, which share similar characteristics at different temperatures. The saturation magnetization reduces with doping Mn and/or Ti contents. Mn doping increases coercivity [41] and hinders the formation of noncollinear spin order [33]. Ti incorporation significantly reduces the coercivity by decreasing the uniaxial magnetocrystalline anisotropy along the *c*-axis [19,21,22].

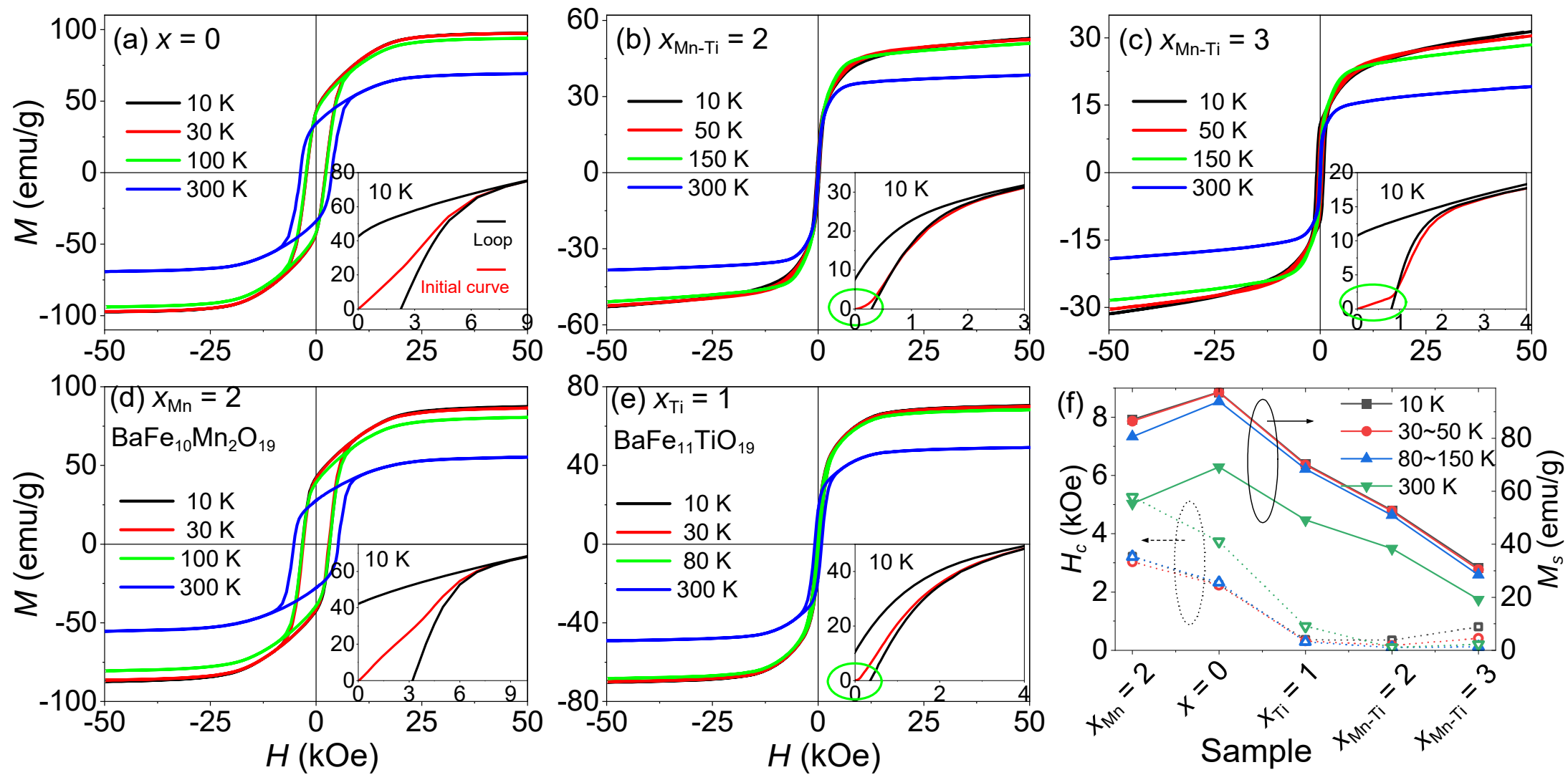

Fig. 6 Hysteresis loops of (a)−(c) $BaFe_{12-2x}Mn_xTi_xO_{19}$ ($x$ = 0, 2, and 3), (d) $BaFe_{10}Mn_2O_{19}$, and (e) $BaFe_{11}TiO_{19}$ at different temperatures. The inset shows the initial magnetization curves and hysteresis loops at 10 K. (f) Saturation magnetization and coercivity of samples at different temperatures.

The magnetic property is governed by the exchange energy for a given spin configuration $\alpha$, expressed as [42,43]

$$E_{ex} = -\frac{1}{2}\sum_{1,i\neq j}^{N} n_i z_{ij} J_{ij} \mathbf{S}_i^{\alpha} \cdot \mathbf{S}_j^{\alpha} \tag{3}$$

where $n_i$ is the number of atoms in the $i$-th sublattice, $z_{ij}$ is the number of nearest neighbors in $j$-th sublattice to $i$-th sublattice, $\mathbf{S}^{\alpha}_i$ and $\mathbf{S}^{\alpha}_j$ are the spin vectors of the ions in the $i$-th and $j$-th sublattices, and $J_{ij}$ is the exchange integrals to determining the exchange energy strength. The magnetic behavior in M-type hexaferrite is primarily controlled by four main superexchange interactions of $Fe^{3+}(12k)$-$O^{2-}$-$Fe^{3+}(4f_1)$, $Fe^{3+}(2b)$-$O^{2-}$-$Fe^{3+}(4f_2)$, $Fe^{3+}(12k)$-$O^{2-}$-$Fe^{3+}(4f_2)$, and $Fe^{3+}(2a)$-$O^{2-}$-$Fe^{3+}(4f_1)$ chains under consideration of $Fe^{3+}$ spins as shown in Fig. 1(b) and 7(a). These competing interactions lead to compensated collinear alignment of $Fe^{3+}$ spins. The ferrimagnetism arises from the net magnetic moment difference between spin-up and spin-down ions across different sublattices in hexaferrite as shown in Fig. 7(b), exhibiting strong uniaxial magnetocrystalline anisotropy along the $c$-axis.

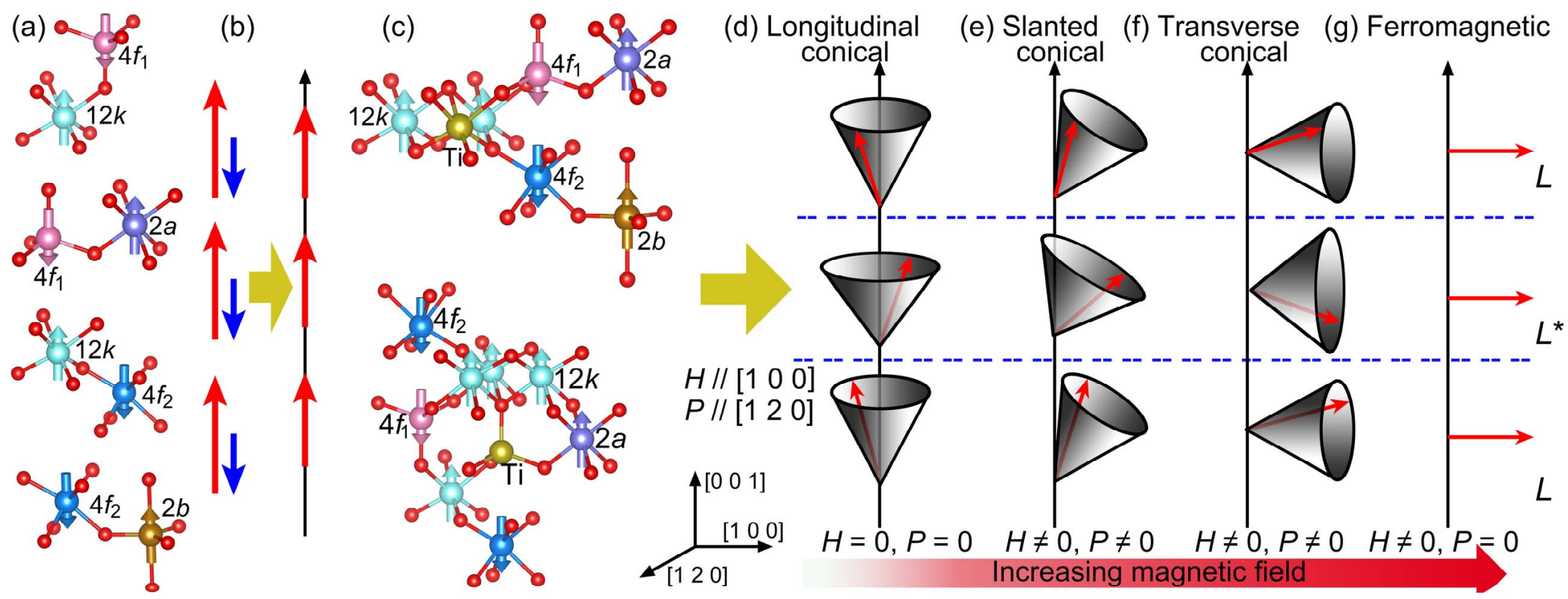

Fig. 7 Schematic illustration of noncollinear longitudinal conical spin-order evolution in M-type hexaferrite. (a) Four main superexchange interactions, (b) ferrimagnetic spin alignment, (c) substitution of $Ti^{4+}$ to $Fe^{3+}$ at 12*k* and 4$f_1$ sites, and (d) – (g) magnetic structure evolution under an external magnetic field in Ti-doped systems, where the cones have two distinct vertex angles.

The magnetic exchange integrals strongly depend on the total $Fe^{3+}$-$O^{2-}$-$Fe^{3+}$ bond distance ($d = d_1 + d_2$) and superexchange $Fe^{3+}$-$O^{2-}$-$Fe^{3+}$ angle $\theta$ [43]. As shown in Table 3, the $Fe^{3+}$(12*k*)-$O^{2-}$-$Fe^{3+}$(4$f_1$) chain exhibits significantly different bond distances in comparison with the other three interactions, resulting in a weaker exchange integral (~5.5 meV) versus the ~8 meV values of the other chains [42].

Table 3 Initial $Fe^{3+}$-$O^{2-}$-$Fe^{3+}$ bond distance and angle of the four superexchange interactions.

| $d_1$ θ $d_2$ | $Fe^{3+}$-$O^{2-}$-$Fe^{3+}$ bond distance $d_1 + d_2 = d$ (Å) | $Fe^{3+}$-$O^{2-}$-$Fe^{3+}$ angle $\theta$ (°) |
|---|---|---|
| $Fe^{3+}$(12*k*)-$O^{2-}$-$Fe^{3+}$(4$f_1$) | 2.101 + 1.906 = **4.007** | **120.85** |
| $Fe^{3+}$(12*k*)-$O^{2-}$-$Fe^{3+}$(4$f_2$) | 1.926 + 1.985 = 3.911 | 127.65 |
| $Fe^{3+}$(2*a*)-$O^{2-}$-$Fe^{3+}$(4$f_1$) | 1.998 + 1.906 = 3.904 | 125.17 |
| $Fe^{3+}$(2*b*)-$O^{2-}$-$Fe^{3+}$(4$f_2$) | 1.866 + 2.071 = 3.937 | 137.93 |

The preferable substitution of nonmagnetic $Ti^{4+}$ ions at 12*k* and 4$f_1$ sites destroys the strong uniaxial magnetocrystalline anisotropy by interrupting the main superexchange coupling in $BaFe_{12}O_{19}$. This involves three superexchange interactions of $Fe^{3+}$(12*k*)-$O^{2-}$-$Fe^{3+}$(4$f_1$), $Fe^{3+}$(12*k*)-$O^{2-}$-$Fe^{3+}$(4$f_2$) and $Fe^{3+}$(2*a*)-$O^{2-}$-$Fe^{3+}$(4$f_1$) and two different exchange integrals. When $Ti^{4+}$ ions replaces one of spin-up $Fe^{3+}$ ions at 12*k* site at the junction position of $L$ and $L^*$ blocks, the superexchange interactions of $Fe^{3+}(12k)$-$O^{2-}$-$Fe^{3+}(4f_1)$ and $Fe^{3+}(12k)$-$O^{2-}$-$Fe^{3+}(4f_2)$ are interrupted to enhance the interactions of $Fe^{3+}(2a)$-$O^{2-}$-$Fe^{3+}(4f_1)$ and $Fe^{3+}(2b)$-$O^{2-}$-$Fe^{3+}(4f_2)$ in the $L^*$ and $L$ blokes as shown in Fig. 7(c). When $Ti^{4+}$ ions substitutes one of spin-down $Fe^{3+}$ ions at $4f_1$ site in $S$ or $S^*$ block, the interactions of $Fe^{3+}(12k)$-$O^{2-}$-$Fe^{3+}(4f_1)$ and $Fe^{3+}(2a)$-$O^{2-}$-$Fe^{3+}(4f_1)$ are interrupted to improve the interactions of $Fe^{3+}(12k)$-$O^{2-}$-$Fe^{3+}(4f_2)$. The $12k$-site $Fe^{3+}$ ions at the junction of $L$ and $L^*$ blocks have a complex effect because they can interact with four $Fe^{3+}$ sites of $12k$, $4f_1$, $4f_2$, and $2a$ geometrically. The substitution of nonmagnetic $Ti^{4+}$ ions to $Fe^{3+}$ ($12k$) ion interrupts the superexchange interaction between different blocks, reducing the magnetocrystalline anisotropy [10,22] and coercivity. These changed interactions disrupt the magnetic balance, causing spin deviations from the $c$-axis with two distinct vertex angles because $Ti^{4+}$ ions interrupt the magnetic superexchange interactions with two different exchange integrals. The magnetic structure of M-type hexaferrite consists of $L$ and $L^*$ blocks of basic unit stacked alternately. The $Fe^{3+}$ ion spins are collinearly aligned inside one basic unit. While the magnetic spins are noncollinearly aligned between the $L$ and $L^*$ blocks stacked alternately along the [0 0 1] crystal direction [4], forming a longitudinal conical spin order with the helix propagation vectors with two different vertex angles along the c-axis as shown in Fig. 7(d). These two helix propagation vectors will respond at different temperatures due to their different interactions.

The cone angle usually increases with decreasing temperature, and consequently, spin projections on the cone bottom balance each other, resulting in a decrease in net magnetization at low temperatures [19,24]. The pronounced magnetization decrease in Ti-containing samples supports the formation of a conical magnetic structure and the increase in cone angle with decreasing temperature. The presence of two distinct cone angles originates from $Ti^{4+}$-induced disruption of superexchange interactions with two different exchange integrals. These bring about the response of noncollinear spin orders at separated temperatures, representing two characteristic inflections in the FC curves as shown in Fig. 5.

### 3.3. Dielectric property

The dielectric property serves as another essential parameter to understand the MD effect. Fig. 8 displays the dielectric dependence of $BaFe_{12-2x}Mn_xTi_xO_{19}$ ($x_{Mn\text{-}Ti}$ = 0, 1, 2, and 3) ceramics on temperature in 10 − 300 K region, exhibiting different characteristics. The permittivity of pure sample decreases continuously with the temperature in 10 − 175 K region without noticeable frequency dispersion, indicating dominant quantum paraelectric behavior at low temperatures [44]. The Pauli repulsion drives the $Fe^{3+}$ ions inside the triangular bipyramid ($FeO_5$) in *R* block to deviate from their symmetric centers of 2*b* sites, generating electric dipoles [3]. However, quantum fluctuations prevent these dipoles from establishing long-range order, resulting in the observed quantum paraelectric characteristics [44,45].

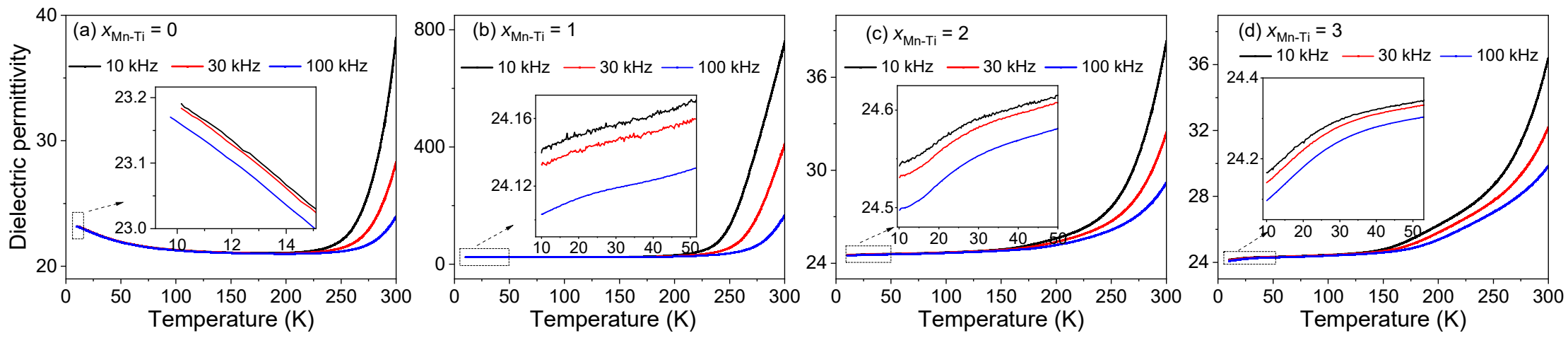

Fig. 8 Temperature dependence of dielectric permittivity of $BaFe_{12-2x}Mn_xTi_xO_{19}$ ($x$ = 0, 1, 2, and 3) ceramics. The insets show the blown-up patterns at low temperatures.

The dielectric parameters of Mn-Ti codoped samples exhibit a little step-like enhancement with temperature in the 10 − 50 K region as shown in Fig. 8(b)–8(d), ascribing to the activated electron hopping accompanied by polaronic effects. The substitution of $Mn^{2+}$ and $Ti^{4+}$ ions significantly modifies the superexchange interactions in Fe-O-Fe networks. The substitution breaks the vibration equilibrium of the $Fe^{3+}$ ions in neighboring triangular bipyramids relative to their central plane to decouple part electric dipoles, which diminishes the quantum paraelectric effect in the low-temperature region. Furthermore, the inevitable oxygen defects in ceramics facilitate the partial reduction of $Fe^{3+}$ to $Fe^{2+}$, generating $Fe^{2+}$-$Fe^{3+}$ defect dipoles. The electron hopping between $Fe^{2+}$ and $Fe^{3+}$ becomes progressively more active with the temperature increase. The test electric field enhances the electron hopping along the $Fe^{2+}$-$Fe^{3+}$ dipole axes, inducing additional polarization that contributes to the dielectric response. Importantly, these defect dipoles interact with the intrinsic dipoles within the triangular bipyramids, enabling electron hopping at a lower temperature [15]. Thus, the increase in dielectric permittivity of Mn-Ti co-doped samples in 10 – 50 K range originates from the electron hopping accompanied by the polaronic effect.

All samples exhibit a rapid dielectric increase accompanied by frequency dispersion in the 175–300 K range, i.e., dielectric relaxation behavior. We measured the variation in dielectric permittivity, dielectric loss (tan δ), and imaginary part of complex electric modulus ($M''$) across a wide frequency range of 1 Hz − 100 kHz to investigate the dielectric mechanism in this temperature region. Fig. 9(a) − 9(d) show that the $BaFe_{12-2x}Mn_xTi_xO_{19}$ samples exhibit two clear dielectric steps in the 125 − 330 K region, accompanying dielectric loss peaks. These features are from two dielectric relaxations, labeled as Dielectric Relaxation I (DR-I) and Dielectric Relaxation II (DR-II).

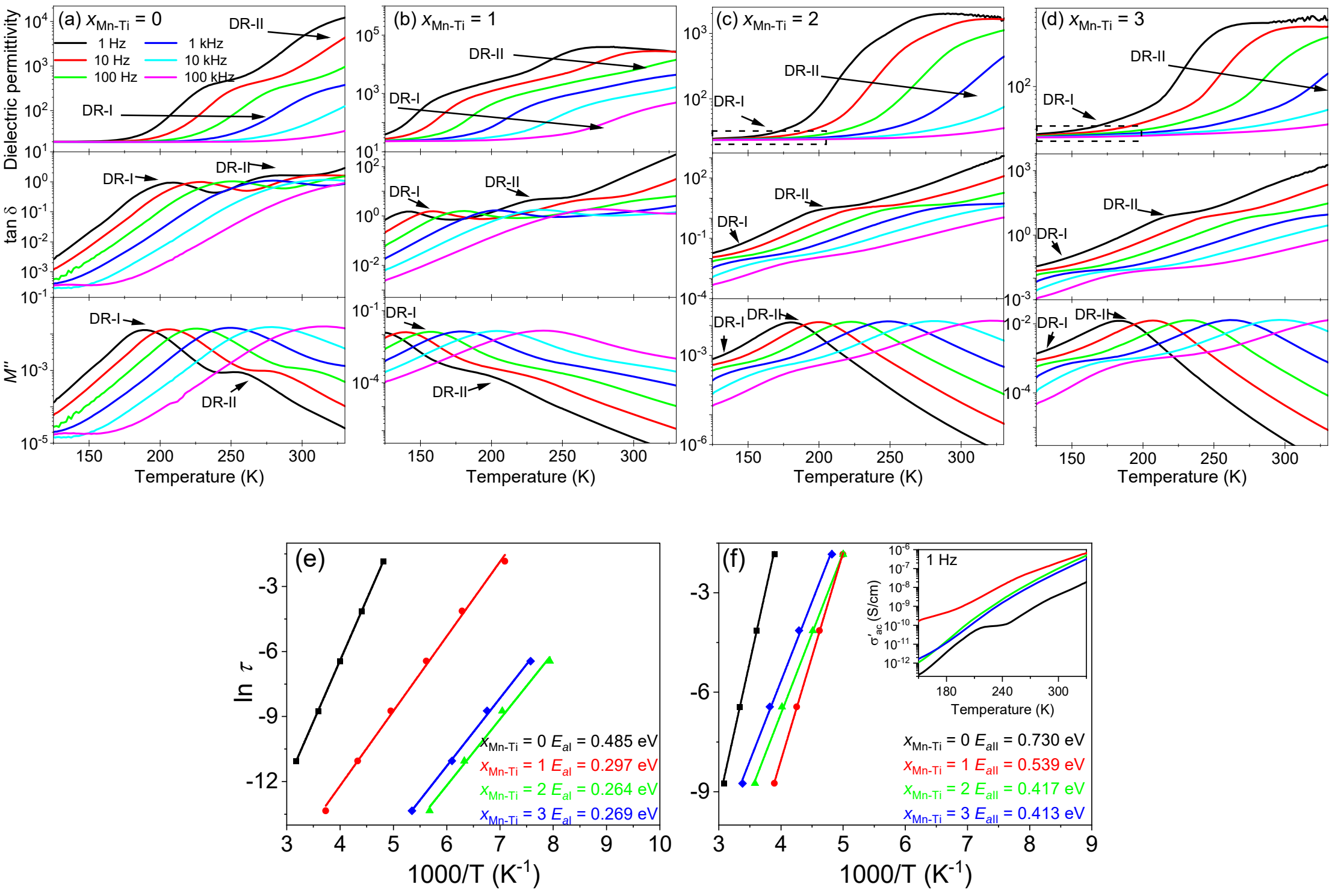

Fig. 9 (a)–(d)Temperature dependence of dielectric permittivity, dielectric loss (tan $\delta$), and imaginary part of the complex electric modulus ($M''$) for the $BaFe_{12-2x}Mn_xTi_xO_{19}$ ($x$ = 0, 1, 2, and 3) ceramics at different frequencies. (e) and (f) Arrhenius fitting for complex electric modulus imaginary part ($M''$) of Dielectric Relaxation I (DR-I) and Dielectric Relaxation II (DR-II). The inset depicts the variation of the real part of ac conductivity ($\sigma'_{ac}$) at 1 Hz with temperature.

The activation energy can be obtained by the Arrhenius law from the electric modulus ($M^* = 1/\varepsilon^*$) [46]

$$\tau = \tau_0 \exp(E_a/k_B T) \quad (4)$$

where $\tau = 1/2\pi f$ is the relaxation time, $f$ is the peak frequency of $M''$, $\tau_0$ is the prefactor, $E_a$ is the activation energy of dielectric relaxation, and $k_\mathrm{B}$ is the Boltzmann constant. Fig. 9(e) and 9(f) display the activation energies $E_{a\mathrm{I}}$ and $E_{a\mathrm{II}}$ obtained from DR-I and DR-II, respectively. The activation energy $E_{a\mathrm{I}}$ corresponding DR-I matches the typical energy of electron hopping between $Fe^{2+}$ and $Fe^{3+}$ in M-type hexaferrite [47,48]. These hopping electrons align along the electric field direction, forming ordered defect dipoles that contribute to dielectric permittivity. This process exhibits relaxation behavior, exhibiting the peaks of tan $\delta$ and $M''$ shift to higher temperatures with the increase in frequency. The hopping of charges becomes easier with increasing temperature due to thermal activation, contributing to dielectric enhancement. The DR-II dielectric relaxation with activation energy $E_{a\mathrm{II}}$ at high temperature results from the Maxwell-Wagner interface polarization in polycrystalline ceramics [8,49,50]. The activated charge carriers migrate in grains with low resistivity but accumulate at grain boundaries with relatively high resistivity under an alternating electric field, generating an interfacial polarization contributing to the frequency-dependent dielectric permittivity at low frequencies [51]. Both DR-I and DR-II intensify gradually with temperature, elevating the baseline value of tan $\delta$, which correlates with the increase in the real part of ac conductivity ($\sigma'_{\mathrm{ac}}$) as shown in the inset in Fig. 9(f). The charge carriers at the grain boundary increase sharply with the temperature to enhance the conductivity exponentially, further confirming the DR-II peak is from the Maxwell-Wagner interface polarization.

The dielectric relaxation mechanism is clearly demonstrated by the Cole-Cole plots of hexaferrite ceramics as shown in Fig. 10(a). These plots exhibit a single arc at low temperature but two separate arcs at high temperature. The resistivities determined from the semicircle intercepts with the $Z'$ axis are summarized in Fig. 10(b), revealing an exponential decrease with temperature. At low temperatures, electrons remain strongly localized around atomic sites, resulting in uniformly high resistivity for both grain and grain boundaries. This uniform behavior manifests as a single high-resistivity arc in the Cole-Cole representation. The electron hopping will be thermally activated with increasing temperature to generate defect dipoles ($Fe^{2+}$-$Fe^{3+}$) in the M-type hexaferrite ceramics [52], leading to an exponential reduction in grain resistivity. Consequently, a pronounced resistivity contrast develops between the conductive grain interiors and resistive grain boundaries. This differentiation produces an additional low-frequency arc in the Cole-Cole plots, corresponding to the grain contribution with the DR-II relaxation observed in Fig. 9(a)–9(d).

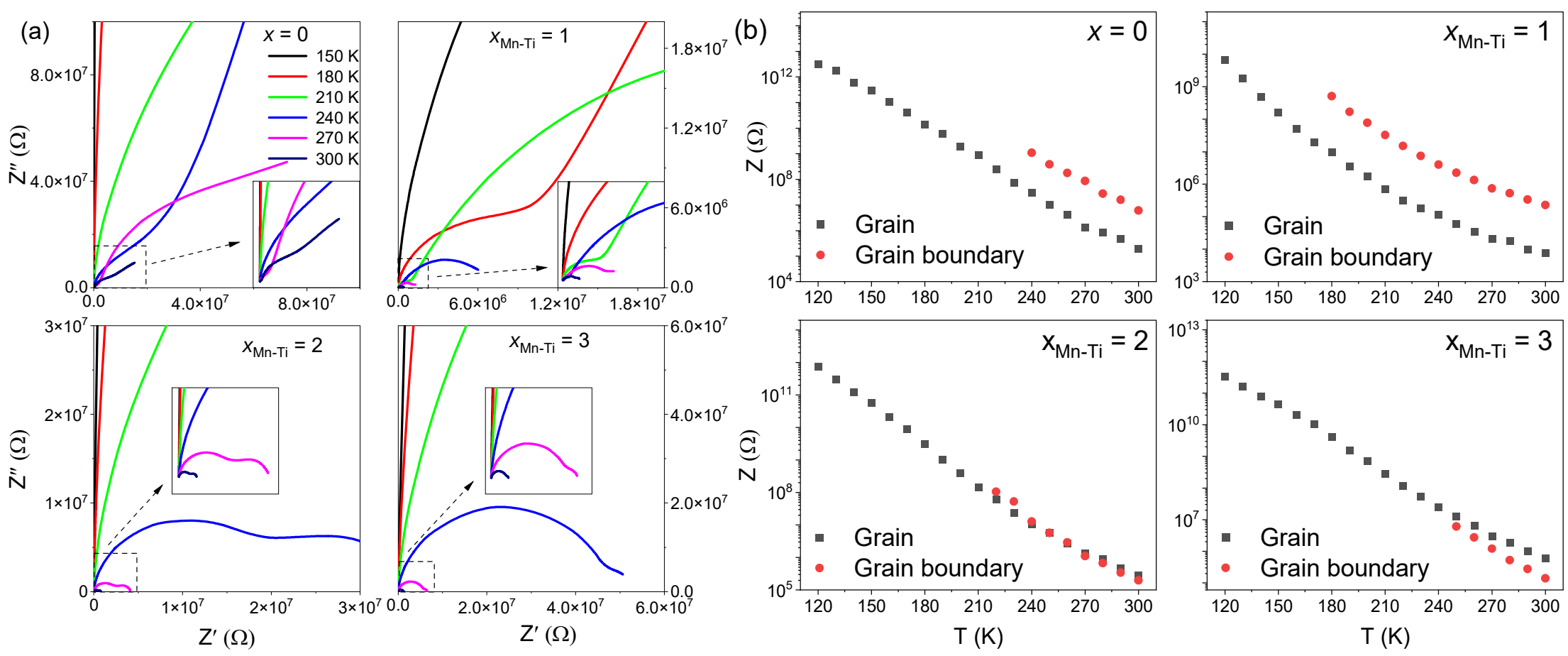

Fig. 10 (a) Impedance Cole-Cole plots of $BaFe_{12-2x}Mn_xTi_xO_{19}$ ($x$ = 0, 1, 2, and 3) ceramics at some selected temperatures. The inset shows a partial enlargement. (b) Impedance of grain and grain boundary.

### 3.4. Magnetodielectric effect

The MD coefficient was calculated with the following equation [8,40]

$$\frac{\Delta\varepsilon_r'}{\varepsilon_r'(50\ \text{kOe})} = \frac{\varepsilon_r'(H) - \varepsilon_r'(50\ \text{kOe})}{\varepsilon_r'(50\ \text{kOe})} \quad (5)$$

where $\varepsilon_r'(H)$ and $\varepsilon_r'(50\ \text{kOe})$ are the dielectric permittivity at 100 kHz in a varying magnetic field and a maximum magnetic field of 50 kOe, respectively. The samples were first cooled down to 10 K in a zero magnetic field environment for the MD measurement. The blue curve corresponds to the field sweep from +50 kOe to −50 kOe, and the red curve corresponds to the magnetic field increase from −50 kOe to +50 kOe.

After completing the measurements at 10 K, the magnetic field was removed and the sample was heated to the next target. This process was repeated sequentially up to 300 K to investigate the temperature-dependent MD response.

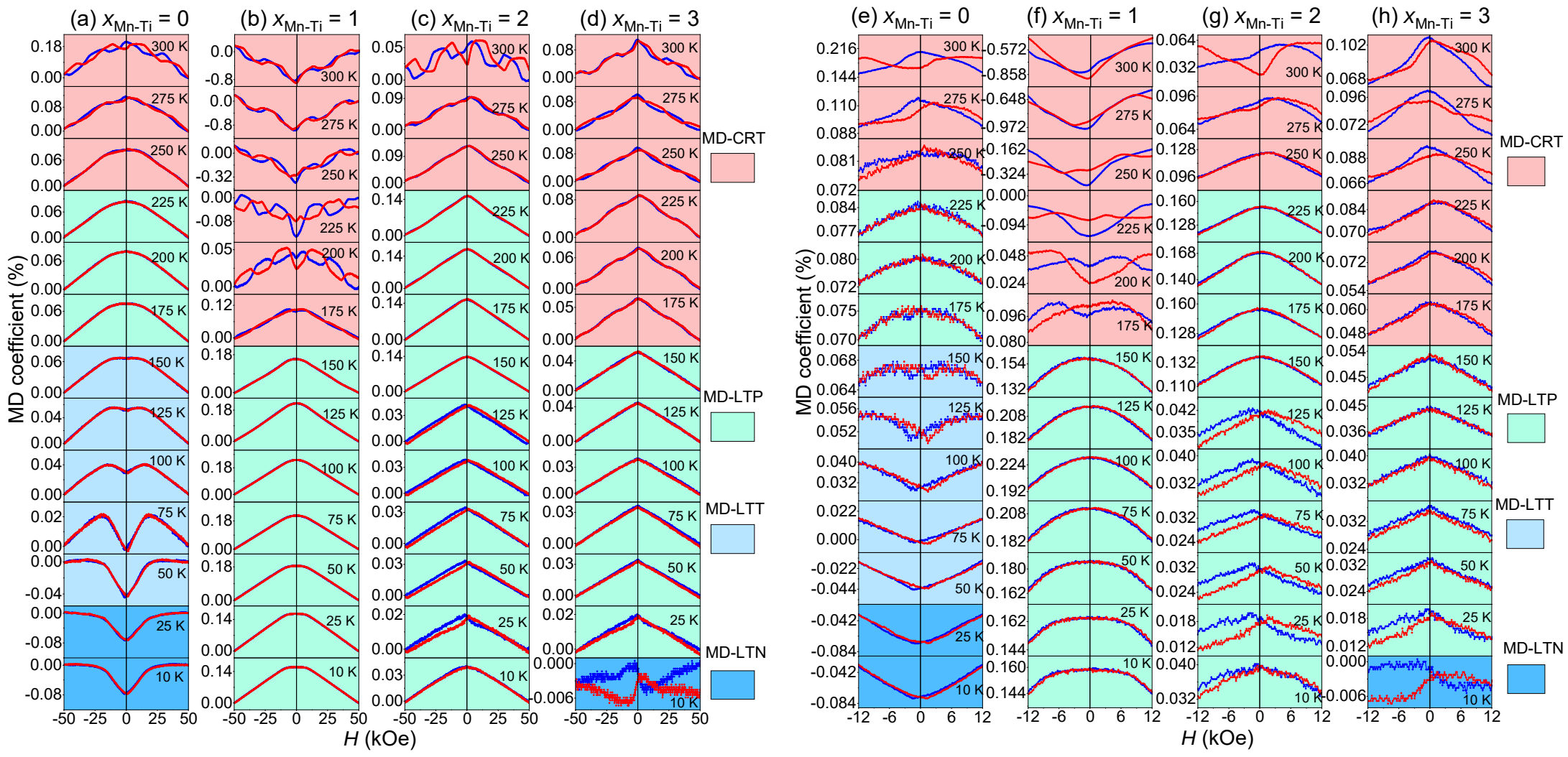

Fig. 11 (a)–(d) Magnetic field dependence of magnetodielectric (MD) coefficient at 100 kHz for $BaFe_{12-2x}Mn_xTi_xO_{19}$ ($x$ = 0, 1, 2, and 3) ceramics at different temperatures. (e)–(h) Their blown-up patterns at low magnetic field. The blue curves show the MD change from +50 kOe to −50 kOe, while the red curves show a reverse change.

Fig. 11 displays the variation of the MD coefficient of $BaFe_{12-2x}Mn_xTi_xO_{19}$ ($x_{Mn-Ti}$ = 0, 1, 2, and 3) ceramics with magnetic field. We identify four distinct temperature regions according to the MD response characteristics. They are the negative MD effect in low temperature (MD-LTN) region, transition MD in low temperature (MD-LTT) region, positive MD effect in low temperature (MD-LTP) region, and MD effect close to room temperature (MD-CRT) region. Both pure $BaFe_{12}O_{19}$ and heavily doped $BaFe_6Mn_3Ti_3O_{19}$ ($x_{Mn-Ti}$ = 3) demonstrate negative MD coefficients in the MD-LTN region. The MD coefficient of the pure sample monotonically increases with magnetic field within ±20 kOe before reaching saturation at higher fields. But $BaFe_6Mn_3Ti_3O_{19}$ exhibits different MD characteristics. The MD coefficient first decreases, then increases near zero field, and lastly decreases at the negative magnetic field during decreasing the magnetic field from +50 to −50 kOe. And the MD coefficient changes reversely during increasing the magnetic field from −50 to +50 kOe. The observed results suggest significant modifications in magnetoelectric coupling induced by Mn-Ti co-doping, which is particularly evident in the low-field region where the MD response shows enhanced sensitivity to magnetic field. The different MD characteristics between the pure $BaFe_{12}O_{19}$ and Mn-Ti codoped samples are attributed to their spin order evolution under applied magnetic fields.

We present the MD curves measured during the initial magnetization stage from 0 to 50 kOe at 10 K, as shown in Fig. 12. The initial MD curve of pure $BaFe_{12}O_{19}$ lies entirely within its MD loop, but the initial MD curves are partially outside the corresponding MD loops for the Mn-Ti co-doped samples, especially for the sample with $x_{\text{Mn-Ti}} = 3$, similar to their hysteresis loops in Fig. 6. The MD effect of pure $BaFe_{12}O_{19}$ in the MD-LTN range results from the spin-phonon coupling, which can modulate the exchange integral under a magnetic field through a phonon renormalization form $\Delta\omega = \lambda\langle \boldsymbol{S}_m \cdot \boldsymbol{S}_n \rangle$ [8,32,53], where $\langle \boldsymbol{S}_m \cdot \boldsymbol{S}_n \rangle$ and $\lambda$ are the correlation interaction and coupling constant, respectively. The spin-phonon coupling plays an important role in mediating the MD effect in the MD-LTN region. This type of MD effect can be described in terms of the Ginzburg-Landau theory.

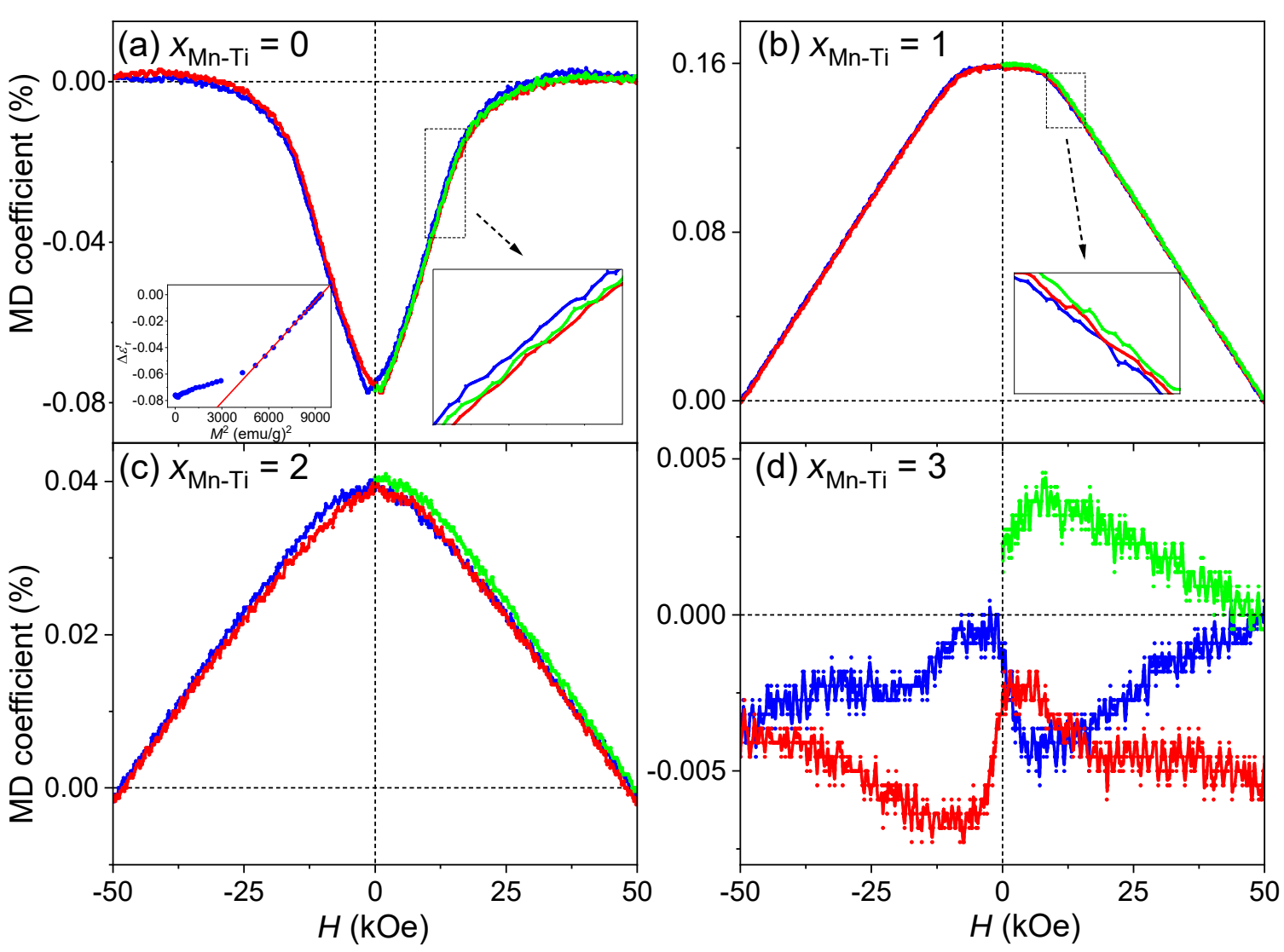

Fig. 12 (a)–(d) MD coefficient of $BaFe_{12-2x}Mn_xTi_xO_{19}$ ($x$ = 0, 1, 2, and 3) ceramics at 10 K with magnetic field. The green curves indicate the initial magnetization stage from 0 to +50 kOe. The blue and red curves indicate the magnetization sweeps from +50 kOe to −50 kOe and from −50 kOe to +50 kOe, respectively. The left inset in (a) shows the square of magnetization ($M^2$) dependence of the dielectric difference ($\Delta\varepsilon_r'$) for the pure $BaFe_{12}O_{19}$ at 10 K.

The free energy ($F$) of the multiferroic system can be expressed as [4]

$$F = \frac{1}{2\varepsilon_0}P^2 - PE - \alpha PM + \beta PM^2 + \gamma P^2 M^2 \qquad (6)$$

where $\varepsilon_0$, $P$, $E$, and $M$ are the vacuum dielectric constant, polarization, electric field, and magnetization, respectively. $\alpha PM$ and $\beta PM^2$ are related to the magnetoelectric effect, $\gamma P^2 M^2$ is related to the MD effect [54]. The dielectric permittivity is the second derivative of free energy in terms of polarization. Then, the MD effect controlled by spin-phonon coupling presents as $\Delta\varepsilon = \partial^2 F / \partial P^2 \propto M^2$ [55,56]. There is a linear relationship between $\Delta\varepsilon'_r$ and $M^2$ in 10 − 35 kOe range as shown in the inset in Fig. 12(a), confirming that the MD effect of pure $BaFe_{12}O_{19}$ in MD-LTN range primarily stems from the spin-phonon coupling. The deviation from linearity in the low-field region (0 − 10 kOe) is ascribed to the magnetization partly contributed by magnetic domain wall movement.

The negative MD effect of Mn-Ti co-doped sample with $x_{\text{Mn-Ti}} = 3$ in the MD-LTN region originates from the spin-induced electric polarization in noncollinear spin order under an applied magnetic field [8,17,24]. The MD curve at the initial magnetization stage lies entirely outside the MD curves in the magnetization stage, which turn around near zero field as shown in Fig. 12(d). These characteristics are similar to its hysteresis loop in Fig. 6(c), suggesting that the spin configuration in initial magnetization differs irreversibly from that in subsequent field cycles. In the MD-LTN region, the noncollinear spin order gradually evolves from a longitudinal conical spin order to a tilted conical spin order, a transverse conical spin order, and finally to a collinear ferrimagnetic spin order with the enhancement of magnetic field, as shown in Fig. 7(d)-7(f). The sample with noncollinear spin order in doped samples produces macroscopic polarization $\boldsymbol{P}$ through the inverse Dzyaloshinskii-Moriya interaction under an applied magnetic field [14]

$$\boldsymbol{P} \propto \sum \boldsymbol{e}_{ij} \times \left(\boldsymbol{S}_i \times \boldsymbol{S}_j\right) \qquad (7)$$

where $\boldsymbol{e}_{ij}$ is the unit vector connecting neighboring spins ($\boldsymbol{S}_i$, $\boldsymbol{S}_j$). The magnetoelectric response is realized by the noncollinear spin order with field dependence, thus realizing the intrinsic MD effect.

The three Mn-Ti co-doped samples share a noncollinear magnetic structure modulated by the Ti element as described in the magnetic section. This magnetic structure should theoretically induce intrinsic electric polarization through noncollinear spin order under a magnetic field. But the samples with $x_{\text{Mn-Ti}}$ = 1 and 2 do not show a negative MD effect at 10 K, which is governed by the electron hopping modulated by a magnetic field. Fig. 13(a) reveals that the negative MD effect of $BaFe_6Mn_3Ti_3O_{19}$ gradually shifts to a positive MD effect with increasing temperature. Fig. 13(b) plots the variation of dielectric permittivity with temperature for the Mn-Ti co-doped samples, demonstrating that the electron hopping occurs at relatively higher temperatures with the increase in doping concentration. These suggest that the absence of negative MD effect for the Mn-Ti co-doped samples with $x_{\text{Mn-Ti}}$ = 1 and 2 stems from the electron hopping modulated by a magnetic field. Thus, these two samples may exhibit a negative MD effect below 10 K, where the contribution from noncollinear spin order becomes dominant.

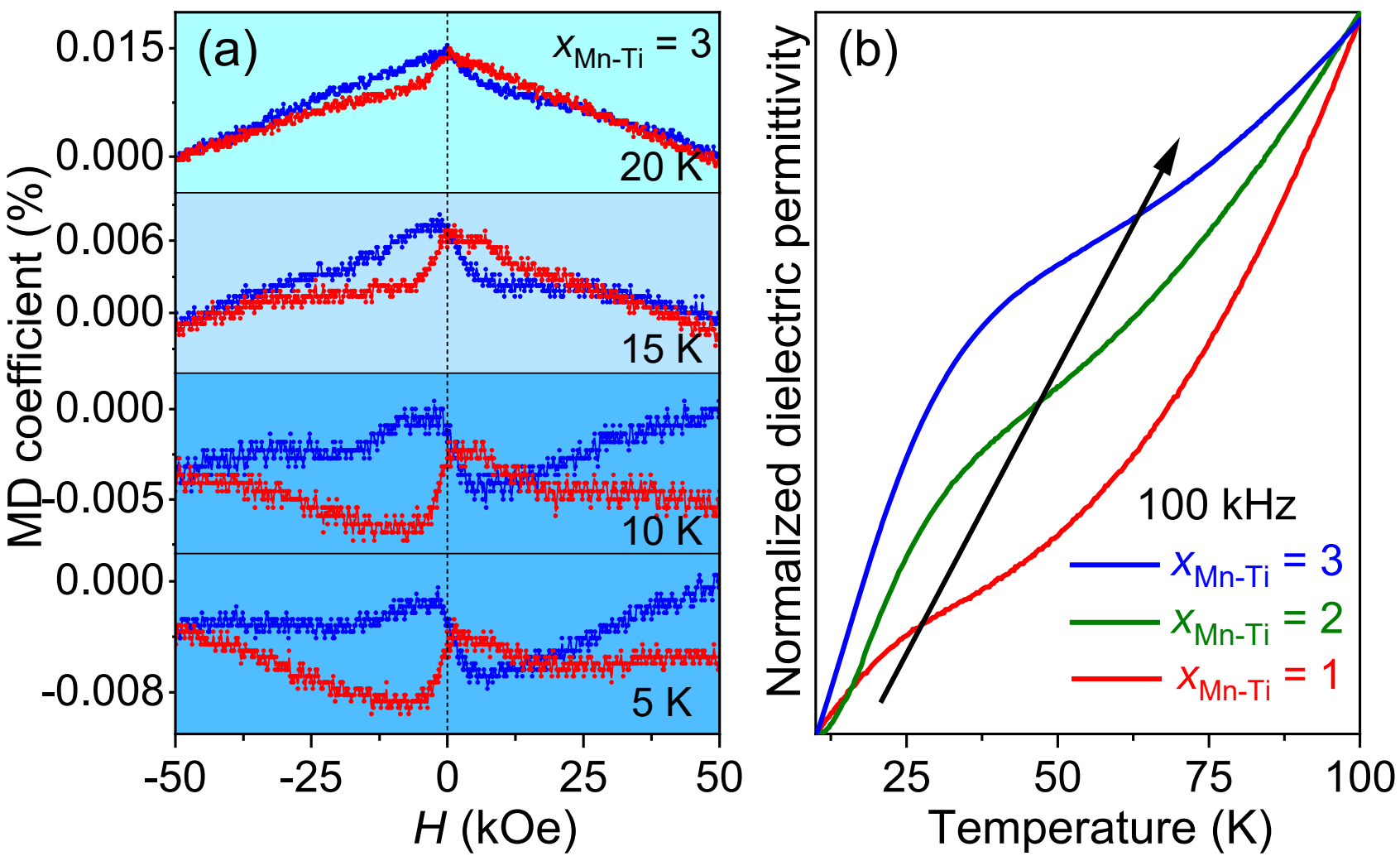

Fig. 13 (a) Variation of magnetodielectric coefficients with magnetic field at 5 K, 10 K, 15 K, and 20 K for $BaFe_6Mn_3Ti_3O_{19}$ ceramic sample, and (b) variation of normalized dielectric permittivity with temperature at 100 kHz for the $BaFe_{12-2x}Mn_xTi_xO_{19}$ ($x$ = 1, 2, and 3) ceramics.

The MD coefficient of pure sample transitions from negative to positive in the MD-LTT region, which is the intermediate stage between the MD-LTN and MD-LTP regions. In the MD-LTP region, all samples exhibit positive MD coefficients that decrease with increasing magnetic field. As discussed in the dielectric section, the DR-I dielectric relaxation originates from the electron hopping inside the material. Under an applied magnetic field, the Lorentz force impacts this electron hopping to modify the dielectric response. Thus, the MD effect in this region primarily originates from the modulation of the electron hopping by a magnetic field.

In the MD-CRT region, the MD effect originates from a combination of magnetoresistance and Maxwell-Wagner interfacial polarization [9], because the Maxwell-Wagner interfacial polarization plays a dominant role in the DR-II dielectric relaxation as shown in Fig. 9. The dielectric contribution from the Maxwell-Wagner interfacial polarization depends on the grain boundary and bulk resistivity of the ceramics. The external magnetic field changes the resistivity of the M-type hexaferrite through the magnetoresistance, consequently changing its dielectric permittivity to achieve the extrinsic MD effect. But the magnetoresistance is weak in the *M*-type hexaferrite [57], resulting in small and unstable MD coefficients with the magnetic field in MD-CRT range.

## 4. Conclusions

In summary, we systematically investigated the influence of Mn and Ti co-doping on the structure, magnetic, dielectric, and MD properties of M-type barium hexaferrite. Mn-Ti co-doping could increase the Ti content in $BaFe_{12}O_{19}$ lattice. $Mn^{2+}$ ions preferentially occupy the $4f_2$ and $2b$ sites while $Ti^{4+}$ ions predominantly substitute $Fe^{3+}$ ions at $4f_1$ and $12k$ sites according to the Raman spectroscopy and first-principle calculation. $Ti^{4+}$ ions play a decisive role in inducing noncollinear spin order, which responds at different temperatures because the $Ti^{4+}$ ions disrupt $Fe^{3+}$-$O^{2-}$-$Fe^{3+}$ superexchange interactions with two main exchange integrals. Pure $BaFe_{12}O_{19}$ maintains ferrimagnetism throughout the $0 - 300$ K temperature range, while the Mn-Ti co-doped $x_{Mn\text{-}Ti}$ = 1, 2 and Ti mono-doped $x_{Ti}$ = 1 samples exhibit a magnetic transition from noncollinear longitudinal conical magnetism to ferromagnetism at 135 K, 227 K, and 45 K, respectively. The emergence of noncollinear magnetism was further corroborated by a significant reduction in coercivity and partial deviation of the initial magnetization curve from the hysteresis loop at 10 K. The Mn-Ti co-doped samples display a step-like dielectric enhancement at low temperatures, attributed to electron hopping accompanied by polaronic effects. All samples exhibit a gradual increase in dielectric permittivity with temperature, accompanied by frequency dispersion in the 175 − 300 K range, resulting from the combined contributions of electron hopping and Maxwell-Wagner interfacial polarization. The MD effects show distinct characteristics across different temperature regions. The negative MD effect of the pure $BaFe_{12}O_{19}$ sample originates from spin-phonon coupling, while that of the Mn-Ti co-doped samples originates from the magnetic-field-modulated noncollinear spin-induced electrode polarization in the MD-LTN region. The positive MD effect in all samples stems from the magnetic field-regulated electron hopping in the MD-LTP region. Whereas, the MD response in the MD-CRT region is predominantly an extrinsic phenomenon governed by the combination of Maxwell-Wagner interfacial polarization and magnetoresistance. These results suggest that Mn-Ti co-doping can promote the hexaferrite MD responses at low magnetic fields through the selective replacement of $Fe^{3+}$ ions, offering new insights for designing functional magnetoelectric materials.

## Acknowledgments

This work was supported by the Open Project Program of Key Laboratory of Functional Materials and Devices for Informatics of Anhui Higher Education Institutes, Fuyang Normal University (No. FSKFKT004D), Fundamental Research Funds for the Central Universities (No. GK202402002) and College Students' Innovative Entrepreneurial Training Plan Program (Grant No. S202410718249).